\documentstyle[12pt]{article}
\begin{document}
\baselineskip     24pt
\thispagestyle{empty}

\begin{center}

{\Large\bf
Scaling, Propagation, and Kinetic Roughening of Flame Fronts in Random Media
}
\end{center}

\baselineskip 15pt

\medskip

\begin{center}

{Nikolas Provatas,$^{1}$ Tapio Ala-Nissila,$^{1,2,3}$
 Martin Grant,$^{1}$ K. R. Elder,$^{1}$ \\ and Luc Pich\'e$^{1,4}$}

\bigskip
\bigskip

{ \it
$^1$Physics Department, Rutherford Building, McGill University, \\
3600 rue University, Montr\'eal, Qu\'ebec, Canada H3A 2T8 \\
}

\bigskip

{\it
$^2$Research Institute for Theoretical Physics, University of Helsinki, \\
P. O. Box 9 (Siltavuorenpenger 20 C), FIN--00014 University of Helsinki, \\
Finland, and Department of Physics, Tampere University of Technology, \\
P.O. Box 692, FIN--33101 Tampere, Finland \\
}

\bigskip

{\it
$^3$Department of Physics, Brown University, Box 1843, \\
Providence, R.I. 02912, U. S. A.\\
}

\bigskip

{\it
$^4$Industrial Materials Institute, National Research Council of Canada, \\
75 De Mortagne Boulevard, Boucherville, Qu\'ebec, Canada, J4B 6Y4 \\
}

\end{center}

\vfill

We introduce a model of two coupled
reaction--diffusion equations to describe
the dynamics and propagation of flame fronts in random media.
The model incorporates heat diffusion, its dissipation,
and its production through coupling to the background reactant density.
We first show analytically and numerically that there is
a finite critical value of the background density, below which
the front associated with the temperature field
stops propagating. The critical exponents associated
with this transition are shown to be consistent with mean field
theory of percolation. Second, we study the kinetic roughening
associated with a moving planar flame front above the critical density.
By numerically calculating the time dependent width and equal time
height correlation function of the front, we
demonstrate that the roughening process belongs to the
universality class of the Kardar--Parisi--Zhang interface equation.
Finally, we show how this interface equation can be analytically
derived from our model in the limit of almost uniform background
density.

\vfill

\baselineskip 24pt

\newpage

\section{Introduction}

Systems out of equilibrium undergoing a transition from a
metastable or unstable state to a stable phase often
develop a front or interface between the two phases.  This
front can play a crucial role in determining the
dynamics of the transition.
There are many systems, arising in various areas of
science, that are characterized by the
emergence of such a front, such as domain walls in
the kinetics of phase transitions \cite{Kru91}, and systems undergoing
chemical reactions \cite{Kap91}.
This paper will study
a common but a
particularly spectacular
example involving the emergence of a reaction front in
combustion, where a flame front forms and can propagate
in a medium of randomly distributed reactants
\cite{Wil85}.

A popular approach to the description of moving fronts
involves using
discrete cellular automaton models
or coupled map lattices.  Examples include wave propagation
in excitable media \cite{Che88} and front propagation for the
description of the spread of epidemics \cite{Mur89}.
Despite their superficial simplicity, such lattice models
can exhibit complex behavior. A good example is a recently
studied coupled map lattice with oscillatory local
elements which has been shown to exhibit a wide variety
of complex dynamics \cite{Kap94}. In particular, the
moving front associated with the model was shown to exhibit
kinetic roughening analogous to pure interface growth equations
\cite{Kru91}.

On a more microscopic level, an approach based on continuum
reaction--diffusion equations has been extensively used in
the chemical literature \cite{Kap91}. Often such nonlinear partial
differential equations can be studied from the nonlinear dynamics point
of view to reproduce many experimentally observed phenomena, such as
spiral waves and chemical oscillations \cite{Roux83}.  In the area of
combustion
of laminar flames in continuous media,
the work of Sivashinsky \cite{Siva77,Frank82}
demonstrates how such equations can be qualitatively
mapped into nonlinear interface equations
describing the propagation of flame fronts. Despite much work, however,
many properties of such equations and their connection to interface
growth equations remain poorly understood, as does their precise
quantitative relationship to combustion.

Lattice
models similar to those studied in the papers above have
also been used to study
``forest fire'' models \cite{Alb86,Bak90,Dro92}.  However,
no particular attention has been focused on the properties
of the reaction front.  Indeed, the majority  of automaton
models of forest fires do not include
flame fronts.

In this work our aim is to systematically study the dynamics
of slow combustion by deriving --- from the microscopic physical
principles behind combustion --- a phase--field reaction--diffusion
model \cite{Landau59}.
Some of these results have been given in a short paper \cite{Pro94}.
This model includes, in a realistic manner, the diffusion
of heat, as well as the dissipation and production of heat through
an activated chemical reaction occurring within the background
density field.
While our model also incorporates the effect of convection, this paper
will focus on combustion in the absence of it.  The model is
investigated both analytically and numerically, using
methods developed in the study of phase transitions
to unravel the asymptotic behavior of a self--sustaining combustion
front growing within a medium of
{\it randomly distributed} reactants.
We examine the dynamics of the front from
the percolation point of view, as well as that of kinetic
roughening of interfaces.
Both the formation of the front, as well as its universal dependence
on length and time are examined.  Most importantly, we
show that the combustion front exists only when the reactant
concentration is greater than a critical value $c^* > 0$
in two dimensions. Moreover, we estimate the
scaling exponents associated with the disappearance of the
propagating front, and show that the behavior near
$c^*$ is consistent with that of
a mean--field percolation transition.  Above the critical
concentration, we find
that the combustion front exhibits kinetic
roughening. We then show both analytically and numerically
that the kinetic roughening is described by the nonlinear
Kardar--Parisi--Zhang (KPZ) interface equation
\cite{Kar86}.

In order to be able to refer to a specific example when dealing
with flame propagation we will motivate it
below in the context of forest fires.
The physics associated with forest fires has recently received
increasing attention \cite{Alb86,Bak90,Dro92}
due to the potential relationship
to the concept of self--organized criticality,
introduced by Bak \cite{Bak87} and collaborators.  In most cases
studied to date, forest fires have been modeled through the use
of cellular automaton models on a lattice
\cite{Bak90,Dro92}.  In these works a
collection of trees which can burn and subsequently reappear is
considered.  In contrast, this work focuses on systems
in which the reacting element cannot spontaneously reappear and
the reaction front is defined only as long as there is reactant
present.  Most importantly, our model is constructed from the
fundamental
physics of reaction--diffusion systems, rather than by introducing
lattice rules. Thus it realistically captures the
various physical phenomena associated with reaction fronts.

This paper organized as follows.  In section two we derive our
phase field model.
In section three we
consider the propagation of the flame front, and argue
that there exists a percolation transition in the model,
corresponding to a critical value of the density.  The
nature of this transition is
examined in the mean field limit, and then numerically.
In section four the kinetic roughening of a planar reaction front is
studied.  We first study it numerically,
then we derive an approximate equation of motion for the
interface which is shown to be identical to
the KPZ equation in the long wavelength limit.
Finally, section five
concludes and summarizes the results of this paper.

\section{The Model}

Our model describes flame propagation through the dynamics
of two fields  inherent in the combustion process: the thermal
field and a field describing the concentration of reactants.
Specifically, it consists of two coupled reaction--diffusion
equations, one for the evolution of the thermal field $T(\vec x, t)$
at position $\vec x$ and time $t$, and
the other describing the evolution of the reactant
concentration $C(\vec x , t)$.  This
model realistically incorporates the interplay between thermal
diffusion and local concentration fields.  Within our model,
variations in $T(\vec x,
t)$ are due to three effects:
(i) thermal diffusion through the medium in which the
flames propagate;
(ii) Newtonian cooling due to
coupling to a heat bath; and (iii) generation of heat, limited by
activation, from the reactants.
The second effect, Newtonian cooling,
describes the most simple manner
in which we can incorporate the effect
of a background heat bath fixed at a temperature significantly
lower than the rest of the reaction area \cite{T_o-note}.
While this provides a sensible and physically-motivated method of
coupling reaction and diffusion to a thermal bath, it should be
noted that it may be worthwhile to investigate other
stabilizing mechanisms.
The amount of heat generated in an activated process depends
on the type of combustion system, or more generally reaction
diffusion system one is examining.
We describe the evolution of
the temperature field by

\begin{equation}
\frac{\partial T}{\partial t} = {D} \nabla^2 T
-{\Gamma}[T-T_o]-  \vec V  \cdot {\nabla} T + R(T,C),
\label{add4}
\end{equation}

\noindent where $D$ is the thermal diffusion coefficient,
$\Gamma$ is the thermal
dissipation constant, and $T_o$ is the constant background temperature
of the bath to which the combustion process gives up heat
through Newtonian cooling.  The term $R(T,C)$ is responsible
for chemical activation.
For completeness we
have included convection due to an external source $\vec V$, but we
shall hereafter set this term to zero.

Nonlinearities enter through the reaction
rate $R(T,C)$, which is
limited by the local concentration of
reactants $C(\vec x,t)$ where $C(\vec x,t)$ represents the
local reactant fraction. The specific form of $R(C,T)$ is
dependent on the type of combustion process in question.
Empirically, heat production in any combustion process is given by
the exothermic reaction
\begin{equation}
R(T,C) \propto T^{\alpha} e^{-A/T} C,
\label{gencomb}
\end{equation}
where \cite{Wil85} $\alpha = {\cal O}(1)$ and $A$ is the activation energy
for combustion (Boltzmann's constant has been set to unity).
As the exponential in
Eq. (\ref{gencomb})  must be of order one, the scale of the heat production
is set by the activation energy, as $A^{\alpha}$.  Similarly the time constant
in front of the proportionality sign will dictate the time scale of the
burning process.   It should be noted that while the precise form of
$\alpha$ in Eq. (\ref{gencomb})  sets the energy scale in the problem, the
main dynamics of burning are controlled only by the Arrhenius form $e^{-A/T}$
\cite{Aldushin81}.

Here we use $\alpha =3/2$.  This can be motivated
by a simple model where reactants burn in
steady state with an ideal gas, and
chemical by-products are ignored.  The net effect
of the reaction is to heat the air
surrounding the reactant, elevating it to the (steady state)
temperature of combustion.
The flux of
molecules striking the reacting surface is $N_s=n \sqrt{(T/ 8 \pi m)}$,
where $n$ is the number density of air, and $m$ is the mass per molecule of
air.  Since combustion
is an activated process the probability of a molecule reacting
is proportional to $e^{-A/T}$ where, again, $A$ is
the activation energy of the reaction.
Thus the total number of molecules reacting is $N_r=N_s e^{-A/T}$.
Since the combustion process occurs
in a steady-state with the surrounding air,
the energy flux from the reactant is limited by the local energy flux of
air molecules striking it, given by $3T/2$ per molecule.
Hence the energy $Q$ released per unit reactant area and per unit time
is given by $Q=(3T/2) N_s e^{-A/T}$.  Denoting the typical
reactant area by $a_t$ and the typical volume by $v_t$, the total
energy produced per unit volume and per unit time in a
region of local reactant  concentration
$C(\vec x, t)$ is given by $P_C=(Q a_t / v_t ) C$.  For a cylindrical
reactant geometry where the height is much greater than the radius,
$a_t /v_t = 2/r$.  We then write

\begin{equation}
P_C = (3n/2r)( \sqrt{2/\pi m}) q(T)C,
\label{pc}
\end{equation}

\noindent where

\begin{equation}
q(T) = T^{3/2} e^{-A/T}.
\label{qTdef}
\end{equation}

\noindent Thus the local energy produced per unit time and volume is
proportional to $q(T)$
where the additional factor of $T^{3/2}$ sets the scale of
energy.
Hence, on measuring temperature in units of the activation
energy, we model the reaction rate in Eq.\ (\ref{add4}) as

\begin{equation}
R = \lambda_2 q(T)  C
= -\lambda_1 \frac{\partial C}{\partial t},
\label{2}
\end{equation}

\noindent where $\lambda_1$ is a dimensionless constant.
The constant $\lambda_2$ is simply the prefactor of
$P_C$ divided by $c_p \rho$ and multiplied by $A^{3/2}$,
where $c_p$ is the specific heat and $\rho$ is the mass density of
air.  This gives us the dimensionless temperature change per unit time
corresponding to the heat production $P_C$.  This completes
the formulation, which thereby
gives a rough estimate of the parameters involved in our model.

The main emphasis in the present work
will be for cases where the initial
distribution of the concentration field $C(\vec x,t=0)$ is random,
and where no complete analytic solutions of Eqs. (\ref{add4})--(\ref{2})
are available.

For the remainder of this paper, we consider a two-dimensional
geometry, where a front
initially parallel to the $y$ axis propagates in the $x$ direction.
The dimensionless parameters are set to $D=0.2$, $\Gamma=0.05$, $T_o=0.01$ and
$\lambda_1=8$, and time is measured in units of those for the reaction,
$\lambda_1/\lambda_2$, and length in units of the dimension of the reactant.
In our numerical work, we
initially distribute the reactant randomly such that
at a given lattice site $C(\vec x,t)=1$ with probability $c$ and zero
with probability $1-c$, in which case the average spatial concentration of
reactants is $c$.  The mesh size in space is set to
$\Delta x = 1$, while the mesh size in time is $\Delta t = 0.01$;
tests of smaller mesh sizes give qualitatively similar results.
It is useful to relate these choices of parameters to the specific
example of a forest fire.  For example, the constant
$\lambda_2$  can be found in terms of the density and
specific heat of air and the activation temperature
of wood.  In physical units, we have
$D\sim 1 m^2 s^{-1}$, $\Gamma \sim 0.05 s^{-1}$, $T_o\sim 10K$
and $c_p\sim  5J g^{-1} K^{-1}$ and $A \sim 500 K$.  With the exception
of $T_o$, these are comparable to real systems.  Our small $T_o$ has
been chosen to give enhanced cooling and hence keep diffusion fields
relatively short ranged as compared to the lattice sizes used.
This allows us to perform our numerical
integrations with good accuracy without having to simulate extremely
large systems.  Test runs show that our results are relatively
insensitive to the choice of $T_o$, as one would expect
provided $T_o$ is much less than the reaction energy heat released $A$.

\section{Dynamics of front propagation}

\subsection{Qualitative Features of Flame Fronts}

Before presenting a quantitative analysis of Eq. (\ref{add4}) and (\ref{2}), it
is useful to qualitatively examine the nature of their solutions.
Due to the activated nature of the combustion process, we expect that
a self--sustaining propagating combustion front requires a sufficient
amount of heat to be released during combustion.  The source for this
heat is dependent on the reactant concentration $c$.
Since activation limits
the production of heat, we expect the existence of a
critical concentration $c^*>0$,
below which the fire will spontaneously burn out
due to insufficient heat production.  That is, for $c < c^*$ the
average
velocity of the front is zero, while it is nonzero for higher
concentrations.  For $c > c^*$ the reaction
front can be identified by a single-valued function
which will be used in all quantitative analysis.
We define the local position of the interface, $h(y,t)$,
as the position $x$ where the temperature field
is maximum at a given time and coordinate $y$.
The variable $h(y,t)$ is a single--valued function of
$y$. Thus, the average velocity of the interface is given by
$v(c) = \langle \partial h(y,t)/\partial t \rangle$.

In Figs.\ 1(a) and (b) typical configurations of
the propagating temperature field are shown for $c=0.65$ and 0.225.
Eqs.\ (\ref{add4}) and (\ref{2}) were solved on a lattice using
periodic boundary conditions in the $y$ direction and fixed boundary
conditions in the $x$ direction.  The size of the system is $L$ in
the $y$ direction, while in the $x$ direction it well exceeds of the
total diffusion length of the propagating $T$ field.  The $T$ field
is always contained within the system size due to a tracking program
that continuously follows the flame front, which is moving towards the right.
The dark pixels correspond to
the temperature field, with the highest temperature corresponding to
the darker the pixels. The interface $h(x,t)$ is
outlined by the
darkest pixels.  The light grey pixels to the right of the interface
correspond
to $C(\vec x , 0)=1$.  The fire is started at the far left by
igniting a complete row of ``trees'' at $y=0$. After a short transient,
the propagating fire front assumes a steady--state average velocity
$v(c)$. For lower densities approaching about 0.2,
the front becomes very irregular and finally stops
propagating.  This is in agreement with our qualitative
arguments; below we will present a quantitative analysis of
this phenomenon.

\subsection{Mean Field Theory}

To quantify the behavior of the flame front near
$c^*$, we first examine the mean interface
velocity with particular interest in identifying
the value of $c^*$.  It is
instructive to begin our investigation
with a mean field model.
Consider a uniform distribution of reactants, whose
initial density
variable $C(\vec x,0)$ at every site
is now equal to a constant $c$.
In this description there are no longer
variations in $T$ or $C$ in the $y$ direction.
Assume there exist mean field temperature and
concentration fronts $T_m$, $C_m$ moving with
constant velocity $v_m(c)$.
Using $\partial T/\partial t =- v_m \partial T/\partial x$ and
$\partial C/\partial t = - v_m \partial C/\partial x$ the mean field
model corresponding to Eqs.~(\ref{add4}) and (\ref{2})
can be written as

\begin{equation}
D \frac{\partial^2 T_m}{\partial x^2} + v_m \frac{\partial T_m}{\partial x}
-{\Gamma}[T-T_o] + \lambda_1 C_m q(T_m) =0,
\label{Tmeq}
\end{equation}

\noindent and

\begin{equation}
v_m \frac{\partial C_m}{\partial x}
- q(T_m) C_m =0.
\label{Cmeq}
\end{equation}

We have solved this mean field model numerically, using the same
procedure as described in Sec.\ 3.1, with $C(\vec x, 0)=c$.
We solved for  the mean field front velocity obtaining a
dependence of $v_m( c )$ on $c$ of the form $v_m(c) \propto
(c-c^*)^\phi$, near $c^*=0.19$, where $\phi=0.5$.

The existence of a finite
critical concentration $c^*$ can be also be
seen by examining Eq.\ (\ref{Tmeq}).
Integrating Eq.\ (\ref{Tmeq}) from $- \infty$ to $+ \infty$,
we obtain

\begin{equation}
\int_{-\infty}^{\infty}
\left[ \lambda_1 C_m q(T_m) - \Gamma (T_m - T_o) \right] dx =0.
\label{balance}
\end{equation}

\noindent Equation (\ref{balance}) tells us that in
order to have a steady state,
the energy produced by activation must balance that lost due to thermal
dissipation.
There are two points where the integrand of Eq.\ (\ref{balance})
is identically zero.  The first $x_h$ lies behind $\max(T_m)$
while the second $x_t$ lies ahead of it.
This point $x=x_t$ can thus be defined via

\begin{equation}
\lambda_1 C_m(x_t) q(T_m(x_t)) = \Gamma (T_m(x_t) - T_o),
\label{intersect}
\end{equation}

\noindent where for values of $c$ near $c^*$ we have found that
$C_m(x_t) \approx c$.
Inspection of Eq.\ (\ref{intersect}) shows that
$T_m(x_t)$ increases as $C_m(x_t) \approx c$ decreases. Moreover,
$\max(T_m)$ clearly decreases
as $c$ decreases.  Thus, since
$T_m(x_t) \le \max(T_m)$ there must exist a $c=c^*$
below which Eq.\ (\ref{balance}) no longer holds.

The exponent $\phi=1/2$ in the mean field limit can also
be obtained from the following argument.
Expanding $q(T)$ in Eq.\ (\ref{Tmeq})
around $T_t(c) \equiv T_m(x_t)$ and taking $C_m(x)$ to be a constant,
near $x_t$,
equal to $C_t(c) \equiv C_m(x_t)$,
we find that the leading edge of $T_m$ goes as

\begin{equation}
T_m \sim \exp[- \frac{v_m + \sqrt{v_m^2 -
4D(\lambda_1 C_t q^{\prime}(T_t) - \Gamma) } }{2D} x].
\label{mean}
\end{equation}

\noindent By imposing the requirement that the
leading edge does not develop any
oscillatory components, we obtain the condition

\begin{equation}
v_m \ge \sqrt{4D(\lambda_1 C_t q^{\prime}(T_t) - \Gamma)}.
\label{mean2}
\end{equation}

\noindent Assuming analytic behavior of $C_t(c)$ and
$T_t(c)$ near $c^*$, we write them
as

\begin{equation}
T_t(c) = T_t(c^*) + \frac{d T_t(c^*)}{dc} (c-c^*)
\label{mean3}
\end{equation}

\noindent and

\begin{equation}
C_t(c) = C_t(c^*) + \frac{d C_t(c^*)}{dc} (c-c^*).
\label{mean4}
\end{equation}

\noindent In regions ahead of the temperature field
where thermal dissipation exceeds thermal activation ($x \ge x_t$),
we have already noted that
the concentration profile $C_m$ will not change from its original value $c$.
Thus, around $x=x_t$ we can approximate $C_t(c) \approx c$.
Using this approximation
in Eq.\ (\ref{mean4}) which along with Eq.\ (\ref{mean3}) is substituted into
Eq.\ (\ref{mean2}), we obtain

\begin{equation}
v_m \ge  A ( c - c^* )^{1/2}
\label{vmc}
\end{equation}

\noindent which yields $\phi=1/2$ and implicitly defines $c^*$ through
$c^* = \Gamma / ( \lambda_1  q^{\prime} (T_{t}(c^*)) )$.
The constant
\begin{equation}
A=\sqrt{4D\lambda_1 \Big(
q^{\prime}(T_t(c^*))+c^* q^{\prime \prime}(T_t(c^*))
\frac{dT_t(c^*)}{dc} \Big) }.
\label{mean5}
\end{equation}
Although we could have  expanded the $q(T)$ term
of Eq.\ (\ref{Tmeq}) about  any point,
choosing $x_t$ gives the maximum lower bound in Eq.\ (\ref{mean2}).
This result is also supported numerically.
This analysis leading to Eq.\ (\ref{vmc}) is
analogous to that used in Ref. \cite{Mur89} to
find front velocities in the context of epidemic models.
Near $c^*$  we expect $v(c)$ to attain its lower bound  \cite{Mur89},
according to Eq. (\ref{vmc}).

\subsection{Numerical Results for Front Propagation}

Eqs.\ (\ref{add4}) and (\ref{2}) were numerically
solved on a lattice under the conditions
described above, with a uniform random distribution of density
with $\langle C(\vec x,t=0)\rangle = c$.
We found that for large concentrations,
the mean interface velocity $v(c)$ is again constant after an initial
transient, and increases with $c$.  The transient increases
as $c^*$ is approached.  As in the mean field case
we expect that in the vicinity of $c^*$,
the asymptotic velocity is defined by the relationship
$v(c) \sim (c-c^*)^\phi$, where $\phi$ is a scaling exponent.
Specifically, the numerical determination $v(c)$ in the
case of a random background gave
$c^*=0.19 \pm 0.02$ and $\phi=0.46 \pm 0.09$ for a system $L=200$.
The scaling of $v(c)$ in the case of a random initial  distribution
of reactants is shown in Fig.\ 2.

To incorporate finite--size effects in a systematic
fashion, we use the scaling form

\begin{equation}
v(c,L) \sim L^{-\phi/ \nu } \Omega [(c-c^*) L^{1/ \nu} ].
\label{5}
\end{equation}

\noindent This is the same as
that used in percolation theory \cite{Sta85}.  Here $\nu$
is the correlation length exponent $\xi \sim (c-c^*)^{-\nu}$, and the
scaling function $\Omega(x \rightarrow \infty) \sim x^\phi$.  We
note that we can relate $\phi$ to the percolation transition exponents
through $v(c) \sim \xi/\tau \sim (c-c^*)^{\Delta - \nu}
\equiv (c-c^*)^{\phi}$, where
$\Delta$ is the critical slowing down exponent. In
Fig.\ 3, we show numerical results for $\ln(v(c,L) L^{\phi /\nu}$) vs.\
$\ln((c-c^*) L^{1/ \nu})$ for nine different system sizes ranging from $L=4$
to $L=200$.  Using $c^*=0.19$ and $\phi=0.46$, we find that the best
collapse occurs for $\nu=0.6 \pm 0.1$.

It is striking that the
results for the critical exponents obtained here are consistent
with the mean field theory of percolation, for which $\Delta=
2\phi=2\nu=1$
\cite{Sta79}.  Qualitatively, heat propagation in our model is limited
by a percolation lattice, provided by the random density field $c$.
Below $c^*$, the connected cluster available for front propagation
breaks down, and the fire spontaneously dies out. The mean field nature
of the critical exponents is due to the relatively long
range nature of the diffusion field associated with $T$, as
compared to the typical front widths for the system sizes
studied here.

\section{Kinetic Roughening of the Flame Front}

\subsection{Numerical Results for the Front Roughening}

For $c > c^*$, it is clear from Fig.\ 1 that the propagating interface
associated with $T$ develops large fluctuations and appears rough.  We
can characterize the interface by defining its width through
$w=\langle (h-\langle h\rangle )^2
\rangle^{1/2}$.  Rough interfaces often satisfy the scaling relation
\cite{Kar86,Fam85}

\begin{equation}
w(t,L) \sim t^\beta f \left( \frac{t}{L^z} \right)
\label{wscaling}
\end{equation}

\noindent for large $L$ and $t$, where $f(x \rightarrow \infty)=x^{-\chi /z}$
and $f(x \rightarrow 0) \sim const.$
with $\chi = z \beta$.
An important example of this is the Kardar--Parisi--Zhang (KPZ) interface
equation \cite{Kar86}, for which the exact and nontrivial exponents are
$\beta=1/3$, $z=3/2$, and $\chi=1/2$, for a one--dimensional
interface (growth front).  In our case,
for any given value of $c > c^*$, we expect the width to obey the
scaling form, i.e., for large $t$ we expect $w \sim t^{\beta}$ in the
limit $t \ll L^z$. We similarly expect that when $t \gg L^z$ the width
will scale with the system size via $w \sim L^{\chi}$.

These scaling forms for the interfacial width
can be derived from a more general
crossover scaling scaling ansatz that
couples time, system size and concentration.
Near $c=c^*$, this scaling form is written as

\begin{equation}
w(c,t,L) = \xi(c) W \left( \frac{t}{\tau(c)},\frac{L}{\xi(c)} \right),
\label{Wscale}
\end{equation}

\noindent where $W(x,y) \rightarrow w_s(x) $ for $x/y^z \ll 1$,
with $w_s(x) \sim x^{\beta}$ for $1 \ll x \ll y^z$,
and $W(x,y) \rightarrow w_L(y) $ for $x/y^z \gg 1$
with $w_L(y) \sim y^{\chi}$ for $y \rightarrow \infty$.
Near the percolation threshold
$\tau(c) \sim (c-c^*)^{-\Delta}$ and $\xi \sim  (c-c^*)^{-\nu}$.
With these forms of $\tau(c)$ and $\xi(c)$,
the scaling function in Eq.\ (\ref{Wscale})
couples $t$, $c$ and $L$ analogously to the way in
which $t$, $c$, and cluster mass
are coupled when describing transport
on percolation clusters \cite{Pan84}.

In the limit $t \ll (\tau(c) / \xi^z(c) ) L^z$
Eq.\ (\ref{Wscale}) reduces to

\begin{equation}
w(c,t)=\xi(c) w_s \left( \frac{t}{\tau(c)} \right),
\label{Lbigscale2}
\end{equation}

\noindent where as $x \gg 1$, $w_s(x ) \sim x^{\beta}$ leading to
$w(t) \sim t^\beta$.
In Fig.\ 4 we show the scaled width $w_s$ plotted
vs.\ the scaled time $t_s=t/\tau$,
for seven different values of $c$ with $L = 200$.  For
this $L$, finite--size effects seem to play no discernible role.  The
inset shows the original data set. A transient time $t_0$ has
been subtracted, which has been determined from the point where $v(c)$
reaches a constant value.  From the fitted $\xi(c)$ and  $\tau(c)$
for the data collapse we cannot accurately estimate $\nu$ and $\Delta$,
although they are again consistent with the mean field values.

 From the scaled data of Fig.\ 4, we can
determine the roughening exponent
$\beta$. The running slope of the data from a $\log w_s$ vs.\  $\log t_s$
plot gives an effective exponent
$\beta(t)$, which is shown in Fig.\ 5.  After
an initial transient the slope clearly tends towards $\beta =  1/3$,
which is the exact KPZ value.  We have also analyzed the data by
calculating the difference $w(bt)-w(t)=A (b^\beta -1) t^\beta$, where
$b$ is a constant (e.g., $b=2$).
 From this method we find $\beta=0.34 \pm
0.04$, which is our best estimate for the exponent.

When  $t \gg (\tau(c) / \xi^z(c) ) L^z$
Eq.\ (\ref{Wscale}) reduces to

\begin{equation}
w(c,L) = \xi(c)  w_L \left( \frac{L}{\xi(c)} \right).
\label{Fwscale}
\end{equation}

\noindent In this limit
the width saturates due to finite--size effects and
thus Eq.\ (\ref{Fwscale}) is independent of time.  In the limit of
large $L$ the saturated width satisfies

\begin{equation}
w(c,L) \sim L^{\chi}.
\end{equation}

\noindent Using system sizes
$L=50$, 76, 100, 150, 200, 300, 400 and 600, we obtain $\chi=0.5\pm 0.1$
for $c=0.5$ and $\chi=0.5\pm 0.3$ for $c=0.85$.  The plots yielding
these values of $\chi$ are shown in Fig.\ 6.  In both cases
the value of $\chi$ is consistent with the
exact KPZ value of $\chi = 1/2$.  Our results for $\beta$ and $\chi$
are therefore in good agreement with those of the KPZ equation
\cite{Kar86,Zha92}.

We also note that a plausible way of expressing the crossover
scaling function, for large $L$ and $t$,
is in the form

\begin{equation}
w(t,c,L) \sim (c-c^*)^{\Delta \beta - \nu} t^\beta
F \left( \frac{t}{(c-c^*)^{z \nu - \Delta} L^z} \right),
\label{wcscaling}
\end{equation}

\noindent where $F(x \rightarrow \infty) \rightarrow x^{- \chi / z}$ and
$F( x \rightarrow 0 ) \rightarrow  const.$, with $z=\chi / \beta$.
Equation (\ref{wcscaling}) gives explicitly the $c$ dependent
generalization of the scaling form of Eq.\ (\ref{wscaling}).
We have not, however, been able to test this particular
ansatz with our present data.

Another quantity that can be used to characterize kinetic
roughening is the equal time height difference correlation function,
which is defined by

\begin{equation}
G(r,t) = \langle ( h(y+r,t) - h(y,t) )^2 \rangle.
\label{corel}
\end{equation}

\noindent Asymptotically $G(r,t)$ satisfies

\begin{equation}
G(r,t) \rightarrow r^{2 \chi}
\label{lim1}
\end{equation}

\noindent for $r \ll t^{1/z}$, while for $r \gg t^{1/z}$ (with $t$ fixed)

\begin{equation}
G(r,t) \rightarrow G(t) \sim t^{2 \beta}.
\label{lim2}
\end{equation}

\noindent While these limits can in principle be used to
extract $\chi$ and $\beta$,
determining the asymptotic limits poses practical problems.
They have been overcome, however, by developing a functional
fitting ansatz for $G(r,t)$ \cite{Ala93}. This ansatz can be used to fit
the whole function $G(r,t)$ and allows the
extraction of estimates for all the scaling exponents $\beta$, $\chi$,
and $z$. We have adapted this method by using the fitting form

\begin{equation}
G_f(r,t) = A(t) \left[ \tanh( B(t)^{1/x} r^{2 \chi_f(t) /x} ) \right]^x
\label{corfit}
\end{equation}

\noindent where $A(t)$, $B(t)$ and $\chi_f(t)$ are fitting parameters,
while $x$ is fixed.  In the limit
$1 \ll L^z \ll t$, $G_f(r,t) = A(t) B(t) r^{2 \chi_f(t)}$, which
allows the estimation of $\chi \approx \chi_f(t)$, as
discussed below. In the
other limit of $r \gg t^{1/z}$, $G_f(r,t) \rightarrow A(t) \sim
t^{2 \beta}$. We have not tried the latter estimate here, however.

In Fig.\ 7 the correlation function is plotted at various times,
for $c=0.5$.  The data is fit to the form of
Eq.\ (\ref{corfit}).  The value of $x$ is first
determined from fitting $G(r,t)$
at one particular time, and is subsequently held fixed for all
other times.  For $c=0.5$, $x=3$ gives the best results.
The inset of Fig.\ 7 shows $\chi_f(t)$, with the solid line
representing the exact KPZ value of $\chi =1/2$.
After a short initial transient, $\chi_f(t)$ becomes
roughly a constant and is consistent with the KPZ value.
In Fig.\ 8 the correlation data for $c=0.85$ is
shown.
For $c=0.85$ we found that $x=4$ fits the data most accurately for
all times.
 From the inset of Fig.\ 8, we see that $\chi_f(t)$  is again consistent
with 1/2.

\subsection{Derivation of the Front Equation}

For $c$ near unity, it is possible to
derive analytically an approximate equation of motion describing
the flame front of our model of Eq.\ (\ref{add4})--(\ref{2}).
We can imagine the temperature field
as  being composed of the steady state mean
field $T_m$ plus a small perturbation $\delta T$
caused by non--uniformities
in the reactant distribution.
This approximation becomes more accurate as
$C(\vec x,0)$ approaches a uniform distribution.
Also, as we take the system
size to infinity, we can systematically average out local
fluctuations along the interface, and
retain an equation governing the long wavelength
dynamics of the interface.

We first introduce a relative coordinate system by
the transformation $x=X(u,s)$ and $y=Y(u,s)$, where
$u(x,y) = const.$ are a set of
planes parallel to the interface, while
$s$ is the arclength along the const.--$u$ plane.
Defining  $u_t(s,t) = \partial u / \partial t$,
we write the equations for $T$ and $C$ in these coordinates as,
\begin{equation}
\frac{\partial T}{\partial t} +
u_t \frac{\partial T}{\partial u} =
D \nabla_{s,u}^2 T
- \Gamma[T-T_0] +
 \lambda_1 q(T) C,
\label{int1}
\end{equation}
\begin{equation}
u_t \frac{\partial C}{\partial u} =
-q(T) C,
\label{int2}
\end{equation}
\noindent where the $ \nabla^2_{u,s}$ operator in Eq.\ (\ref{int1})
is given by
\begin{equation}
\nabla_{u,s}^2 =
\frac{\partial^2}{\partial u^2} +K(s) \frac{\partial}{\partial u}
+\frac{\partial^2}{\partial s^2},
\label{grad2}
\end{equation}
and $K(s)$ is the curvature \cite{Rog89}.
In Eq.\ (\ref{int2}) we assume that the reactant field changes much
faster than the thermal field, thus dropping the $\partial C/\partial t$ term.
This is quite common in reaction-diffusion
systems \cite{Kap83}.

Solving first equation (\ref{int2}) we obtain

\begin{equation}
C=\eta(s,t) \exp \left( \int_u^{\infty} \frac{q(T(z,s,t)}{u_t(s,t)} dz \right)
,
\label{ceq}
\end{equation}

\noindent where $\eta(s,t)$ comes from the
boundary conditions, which demand that $C(\infty,s,t)$ be either zero
or one, with the same distribution as the original $C$ field.
Statistically, $\eta(s,t)$ is a Bernoulli random variable with
$\langle \eta \rangle=c$.  The boundary condition as $u \rightarrow - \infty$
is that $C$ vanishes.  This is satisfied since $u_t=-v$, where $v$
is the normal velocity.

Next we examine the $T$ equation.  We represent the $T$ field as
$T = T_m(u) + \delta T(u,s,t)$ where $T_m$ is the mean field solution.
Substituting this expression for $T$ and the form of $C$
given by Eq.\ (\ref{ceq}), into Eq.\ (\ref{int1}) we obtain, to first
order in $\delta T$,
\begin{eqnarray}
u_t \frac{\partial T_m}{\partial u} =
D \frac{\partial^2 T_m}{\partial u^2} + D K(s) \frac{\partial T_m}{\partial u}
-\Gamma[T_m-T_0] +
\nonumber \\
\lambda_1 \eta(s,t) S(T_m(u),u_t)  +
P \cdot \delta T + \lambda_1 \eta(s,t)
\nonumber \ \times \\
\ \times S(T_m(u),u_t)
\Big( I[\delta T] + \frac{q^{\prime}(T_m) \delta T}{q(T_m)} \Big),
\label{int22}
\end{eqnarray}
where the operator $P$ is given by
\begin{equation}
P =\nabla_{s,u}^2 -
{\partial \over  \partial t} - u_t {\partial \over  \partial u} -
\Gamma,
\label{Pop}
\end{equation}
and the function $S(T_m(u),u_t)$ is defined by
\begin{equation}
S(T_m(u),u_t)= q(T_m(u)) \exp \left( \int_u^{\infty}
\frac{q(T_m(z,s,t)}{u_t(s,t)} dz \right),
\label{Sterm}
\end{equation}
with
\begin{equation}
I[ \delta T ] =\int_u^{\infty}
\frac{q^{\prime}(T_m(z))}{u_t} \delta T(z,s,t) dz .
\label{Integ}
\end{equation}

Next we multiply Eq.\ (\ref{int22}) by $\partial T_m/\partial x$ and
integrate the resulting equation from $(-\epsilon, \infty)$, where
$\epsilon$ is defined as a point to the left of $\max(T_m)$.  In
performing this integration, we are essentially following
the methods for deriving interface equations \cite{Rog89}.  For such a
method, the ``projecting field'' $T_m$ must assume two states
over the range $(-\epsilon , \infty )$.  A simple calculation
shows that the trailing edge of $T_m$ (defined on $x < \max(T_m)$)
goes as $T_m \sim \exp(-\Gamma |x|/v)$ while we can write the the
leading edge (defined on $x > \max(T_m)$) as $T_m \sim  \exp(-v
|x|/D)$. For our $\Gamma$, the field $T_m$ can be approximated by a
constant for a certain distance, $\epsilon$, to the left of
$\max(T_m)$. Conversely, since $v/D \gg \Gamma$, the leading edge of
$T_m$ falls to $T_o$ much faster than the trailing edge.  Thus, over the
range for which the reaction front  is defined, we can treat $T_m$ as a
two--state function.  Performing the projection onto $T_m$, we arrive
at

\begin{eqnarray}
\frac{\partial u}{\partial t} =
D K(s)  - \frac{\Gamma \Lambda}{\sigma} +
\frac{\lambda_1 \eta(s,t)}{\sigma} \times
\nonumber \\
\int_{-\epsilon}^{\infty}
\frac{dT_m}{du} S(T_m(u),u_t) du +
\frac{1}{\sigma}\int_{-\epsilon}^{\infty}
\frac{dT_m}{du} \Big( P \cdot \delta T +
\nonumber \\
\lambda_1 \eta(s,t) S(T_m(u),u_t)
( I[\delta T] + \frac{q^{\prime}(T_m)}{q(T_m)} \delta T ) \Big) du,
\label{int3}
\end{eqnarray}
where the constants $\sigma$ and $\Lambda$ are defined by
\begin{equation}
\sigma=\int_{-\epsilon}^{\infty} \left( \frac{dT_m}{du} \right)^2 du,
\label{sigma}
\end{equation}
and
\begin{equation}
\Lambda= \int_{-\epsilon}^{\infty}
\frac{dT_m}{du} (T_m(u) - T_0 ) du.
\label{lambda}
\end{equation}

Since we are interested in the equation of motion of the reaction
front as the system size $L \rightarrow \infty$ and as the
wavenumber $k \rightarrow 0$, we integrate out the effect of the short
wavelengths by introducing the operator

\begin{equation}
\Omega(f(u,s^{\prime},t)) =
\langle f(u,s^{\prime},t)
\rangle_{(s - \frac{L_B}{2} < s^{\prime} < s +\frac{L_B}{2})},
\label{smooth}
\end{equation}

\noindent where $L_B$ represents the distance over which the function
$f(u,s^{\prime},t)$ is averaged.
This operator smooths over a distance $L_B$ along the interface,
thus eliminating wavenumbers larger than $2 \pi /L_B$.
We now proceed to rewrite Eq.\ (\ref{int3})
with respect to $s^{\prime}$ and apply $\Omega$ to both sides of the equation.

The first four terms in  Eq.\ (\ref{int3}) are simplified by
expressing $u_t$ as

\begin{equation}
u_t(s^{\prime},t) = u_t(s,t) + \delta v(s^{\prime},t),
\label{ueq}
\end{equation}

\noindent where $u_t(s,t)$ is the normal velocity
obtained after averaging over the block
$s - L_B/2 < s^{\prime} < s + L_B/2$.
Applying
$\Omega$ to $u_t(s^{\prime},t)$ and treating $\delta v(s^{\prime},t)$ as a
random variable with zero mean, we arrive
at $\Omega(u_t(s^{\prime},t))=u_t(s,t)$.
In a similar manner the curvature and the constant term in  Eq.\ (\ref{int3})
are just rewritten in terms of $s$ when operated on by $\Omega$.
Using Eq.\ (\ref{ueq}) and expanding $S(T_m(u),u_t)$,
the integrand in the fourth term in  Eq.\ (\ref{int3})  gives

\begin{eqnarray}
\frac{\lambda_1}{\sigma} \eta(s^{\prime},t)
S(T_m(u),u_t(s,t)) \Big(1 - ~~~~~~~~~~~~~~~~~
\nonumber \\
\frac{ \delta v( s^{\prime},t ) }{ u_t(s,t) }
\int_u^{\infty} \frac{ q(T_m(z)) }{ u_t(s,t) } dz \Big).
\label{fourthterm}
\end{eqnarray}

\noindent Now, it is reasonable to assume that the fluctuating variables
$\eta(s^{\prime},t)$ and $\delta v(s^{\prime},t)$
are statistically independent with  $\delta v(s^{\prime},t )$
having zero mean in the limit where $L_B \rightarrow \infty$.  Thus,
applying $\Omega$, the fourth term in Eq.\ (\ref{int3}) becomes

\begin{equation}
\frac{\lambda_1}{\sigma} \hat{\eta}(s,t) \int_{-\epsilon}^{\infty}
\frac{dT_m(u)}{du}S(T_m(u),u_t(s,t)) du
\label{omegafourth}
\end{equation}

\noindent where $\hat{\eta}$ is the noise under the action of $\Omega$.
Assuming, further, that $\delta T$  and
$\delta v(s^{\prime},t)$ are also uncorrelated
with $\delta T$ having zero mean, we can similarly show that the fifth
term involving $\delta T$ goes to zero under the action of $\Omega$.
Thus, after smoothing over $L_B$ where $L_B \rightarrow \infty$, the
equation of motion for the reaction front becomes

\begin{eqnarray}
\frac{\partial u}{\partial t} =
D K(s)  - \frac{\Gamma \Lambda}{\sigma} +
\nonumber \\
\frac{\lambda_1}{\sigma} \hat{\eta}(s,t) \int_{-\epsilon}^{\infty}
\frac{dT_m}{du} S(T_m(u),u_t) du.
\label{averaged}
\end{eqnarray}

To simplify Eq.\ (\ref{averaged}) we note that
the time derivative of the interface  $ \partial h / \partial t$
are related by

\begin{equation}
\frac{\partial h}{\partial t} =
\left(1 + (\partial h / \partial x)^2 \right)^{1/2} v
\label{cosine}
\end{equation}

\noindent where $v=-\partial u / \partial t$ is the normal velocity
to the interface.
Also we write the curvature in terms of the function $h(x,t)$  as
\begin{equation}
K=-
\left(1 + (\partial h / \partial x)^2 \right)^{-3/2}
 \frac{\partial^2 h}{\partial x^2}.
\label{cosin}
\end{equation}

\noindent  Furthermore, let us approximate $u_t $ in the function
$S(T_m(u),u_t)$ appearing in Eq. (\ref{averaged})
by $u_t  \approx -v_m$, where
$v_m$ is the (as yet undetermined) mean interface velocity.
Finally we note that we can write the noise term $\hat{\eta}$ as $\hat{\eta}=
\hat{\mu}+c$,
where $c$ is the average reactant density and $\langle\hat{\mu}\rangle=0$.
Using the definitions of this last paragraph in
Eq. (\ref{averaged}) and writing
$h(x,t)=v_mt+ \hat{\zeta}(x,t)$ we obtain, after expanding to second order in
 derivatives of $h(x,t)$:

\begin{equation}
\frac{\partial \hat{\zeta} }{\partial t} =
D \frac{\partial^2 \hat{\zeta} }{\partial x^2} +
\frac{1}{2\sigma}(\Gamma  \Lambda +\hat{\lambda}_1 c) \Big( \frac{\partial
\hat{\zeta} }{\partial x} \Big)^2 +
\frac{ \hat{\lambda}_1 }{\sigma} \hat{\mu},
\label{KPZ}
\end{equation}

\noindent where we identify $v_m = \Gamma \Lambda / \sigma + \hat{\lambda}_1c
/\sigma$,
and  $\hat{\lambda}_1$ is given by

\begin{equation}
\hat{ \lambda}_1 =-\lambda_1 \int_{-\epsilon}^{\infty} \frac{dT_m(u)}{du} S(T
_m(u),-v_m) du.
\label{lamconst}
\end{equation}
The noise term at the interface, $\hat{\mu}$,
is a Bernoulli distributed
random variable, which by the central limit
theorem becomes
normally distributed as $L_B \rightarrow \infty$.
Thus, the equation of motion we have derived for the flame
front is equivalent to the KPZ equation of
interfacial roughening in the long wavelength limit.

\section{Summary and Discussion}

In this work we have developed a realistic phase-field
model for the dynamics of slow combustion in a randomly distributed medium.
Our model is derived from the first principles of chemical
kinetics, assuming a reactant burns via a steady state
reaction with air.  In addition to chemical activation, our model also
includes thermal diffusion and thermal dissipation.
An important property
of our model is the existence of a continuously extended thermal field,
through the
diffusive coupling of the thermal field to the concentration field
of the reactant.
We define the parameters of our model
based on the
properties of wood, and motivate our combustion problem
in the context of forest fires.

We find a percolation transition at a
critical density $c^*\approx 0.19$, below which the flame front will
spontaneously die. We have analyzed the nature of this transition
showing that the velocity of the
average front position scales as $(c-c^*)^{\phi}$,
where $\phi \approx 1/2$.  We also found a correlation exponent
$\nu \approx 1/2$. Both these values are consistent with
the mean field theory of percolation as well as the mean field limit
of combustion we derived from the full equations.
Through analyzing our mean field model,
the existence of a critical concentration is also found
analytically.

Above $c^*$, we found that the interface associated with the
combustion front displays kinetic roughening.  By an appropriate
application of the scaling theories developed
for percolation theory and kinetic roughening of interfaces,
we have derived a scaling ansatz for the interfacial width
that couples time, concentration
and system size.  This form has been verified numerically
in appropriate limits, and used to estimate the scaling
exponents $\beta$ and $\chi$ independently. Together with
the equal time height--height correlation function, the results
give strong evidence to the fact that the kinetic roughening
of the flame front is in the universality class of the
KPZ equation.  We have also obtained this result analytically,
by deriving an approximate
interface equation which, near $c=1$, is equivalent to the KPZ
equation.

While our model was derived to describe combustion fronts,
it also lends itself to the description of a wider class
of activated  reaction-diffusion problems.
Provided the reaction term contains an Arrhenius factor,
the prefactor $T^{\alpha}$ multiplying the activated form
plays a less important role in determining the properties
of the equation. Thus, we believe that a wide class of
different physical systems, described by equations analogous
to Eq. (1), show behavior of the type described here.
Of course, one of the main ingredients in the present work
has been the introduction of a {\it random} background
density field of reactants, which leads to the kinetic
roughening of the front.

\section*{Acknowledgments}

The Centre for Scientific Computing (CSC) Co.\ of Espoo, Finland has
provided most of the computing resources for this work.  This work has
also been supported by the Academy of Finland, the Natural Sciences and
Engineering research Council of Canada and {\it les  Fonds pour la
Formation de Chercheurs et l'Aide \`a la Recherche de Qu\'ebec\/}.

\noindent {\Large \bf Figure Captions}

\begin{enumerate}

\item
{
The temperature field $T(x,y,t)$
for a moving fire front in a uniform, random forest with
(a) $c=0.65$, and (b) $c=0.225$. The dark pixels correspond to
the temperature field; the higher the temperature
the darker the pixels. The interface $h(x,t)$ is
defined by the curve outlined by the
darkest pixels.  The light grey pixels to the right of the interface
represent reactant (trees).
}

\item
{
Scaling of the interface velocity to the form
$v(c) \sim (c-c^*)^{\phi}$ in the case of a random initial
reactant distribution.
This curve was fit using concentrations up to
$c=0.85$, and $L=200$.   The inset shows data of
$\rangle h\rangle$ vs.\ $t$ for $c=0.21,0.22,0.23,0.24$.}

\item
{
Finite size scaling of $v(c,L)$.  The main figure
shows $\ln [v(c,L) L^{\phi / \nu}]$ vs.\ $\ln[(c-c^*)L^{1 / \nu}]$.
The inset shows the unscaled data for system sizes
$L=4,6,8,24,44,54,64,104$, and 200 from right to left.
Sizes larger than $L=24$ lie almost on the same curve.
All systems contain data for $c$ up to and
including $c=0.85$.
}

\item
{
Crossover scaling function $w_s$ plotted vs.\ $t_s=t / \tau$.
The inset shows the concentration dependent
width $w(c,t)$ for $c=0.4,0.5,0.6,0.7,0.75,0.80$,
and $0.85$ from top to bottom.  The roughness increases with decreasing
density.  A transient time $t_0$ and the corresponding
offset $w_0$ has been subtracted from each $w(c,t)$.
}

\item
{
A log--log plot of the scaling function $w_s$ of Fig.\ 4.
The inset shows the
effective $\beta$ as a function of time.  The straight line represents
$\beta = 1/3$.
}

\item
{
Plots of $\ln(w)$ vs.\ $\ln(L)$ for $c=0.5$ and $0.85$.  The slopes
give, respectively, $\chi=0.5 \pm 0.1$ and $\chi=0.5 \pm 0.3$,
both consistent with
the exact KPZ value of $\chi=1/2$.
}

\item
{
Correlation function $G(r,t)$ and the fitted function
$G_f(r,t)$ (solid line) versus $r$ at different times, for $c=0.5$.
The higher curves represent larger times.  The fitting function
is given in the text.  The  inset shows $\chi_f(t)$
versus time with the straight line representing the exact
KPZ value of $\chi=1/2$.  }

\item
{
Correlation function $G(r,t)$ and the fitted function
$G_f(r,t)$ (solid line) versus $r$, for $c=0.85$. See text for details.
}

\end{enumerate}


\begin{thebibliography}{99}
\bibitem{T_o-note}
Newtonian cooling implies heat can be removed without diffusion.
This stabilizes the constant-velocity motion of the flame.
If, instead, only diffusion was present, we expect a flat front
to be unstable by the Mullins-Sekerka mechanism.
\end{thebibliography}
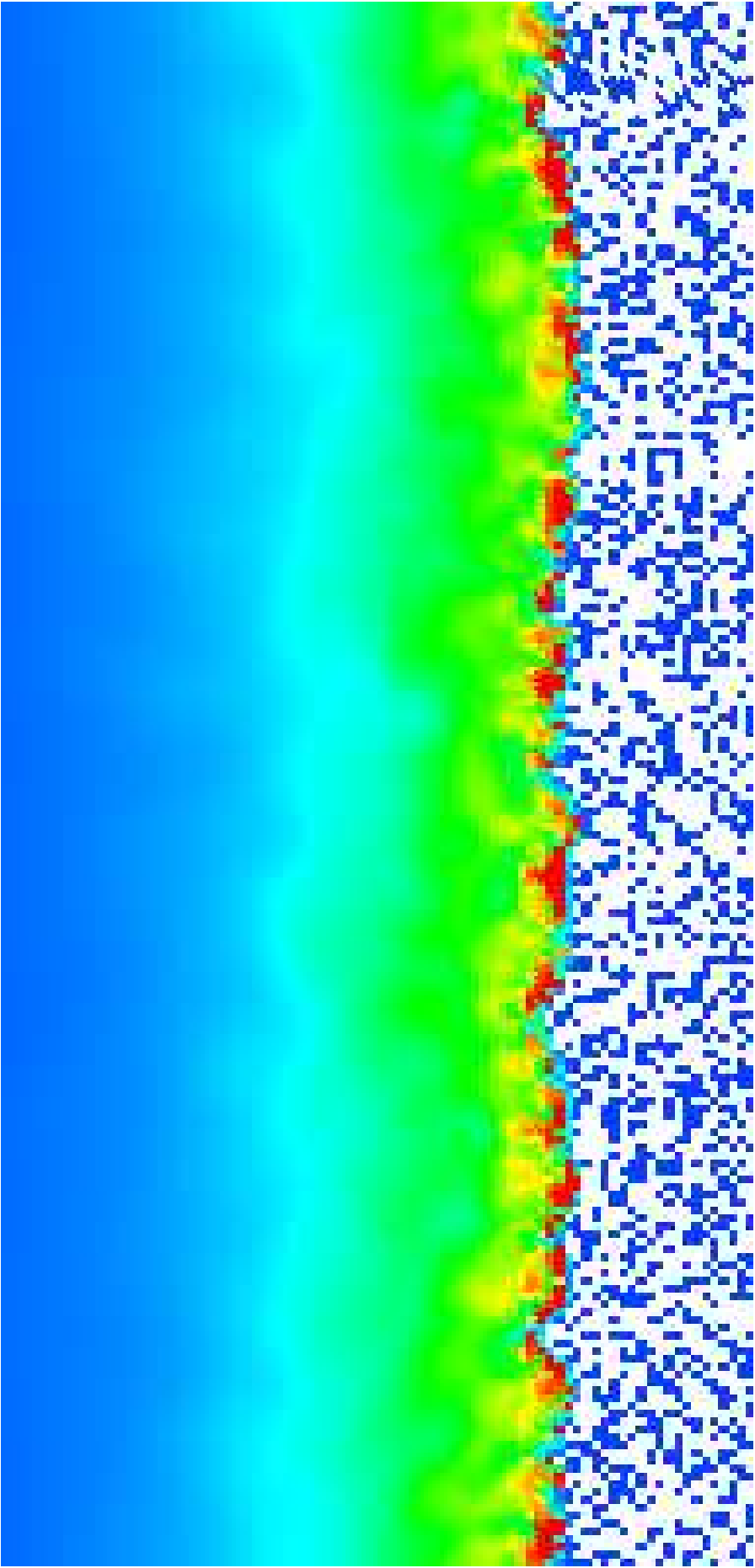
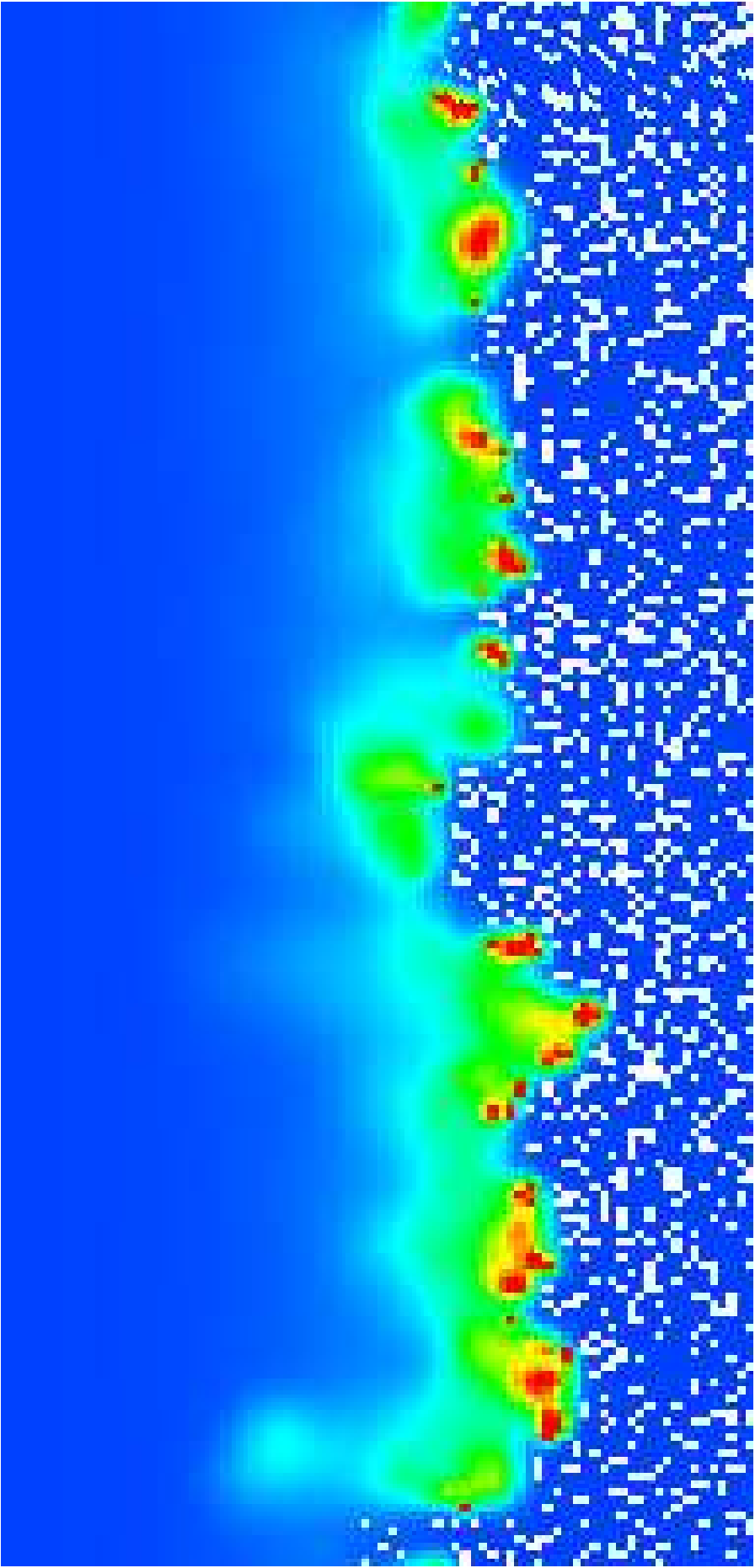
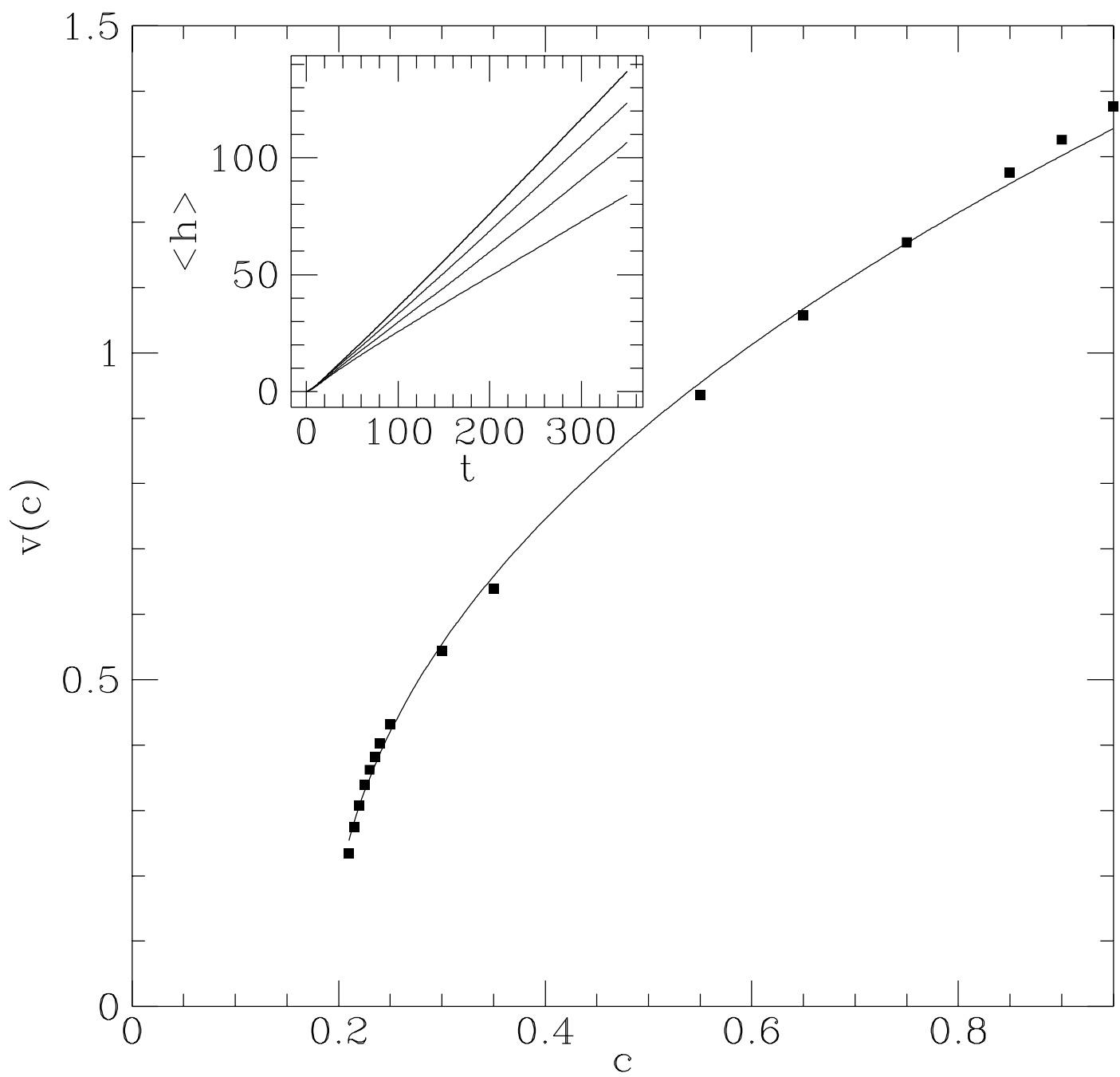
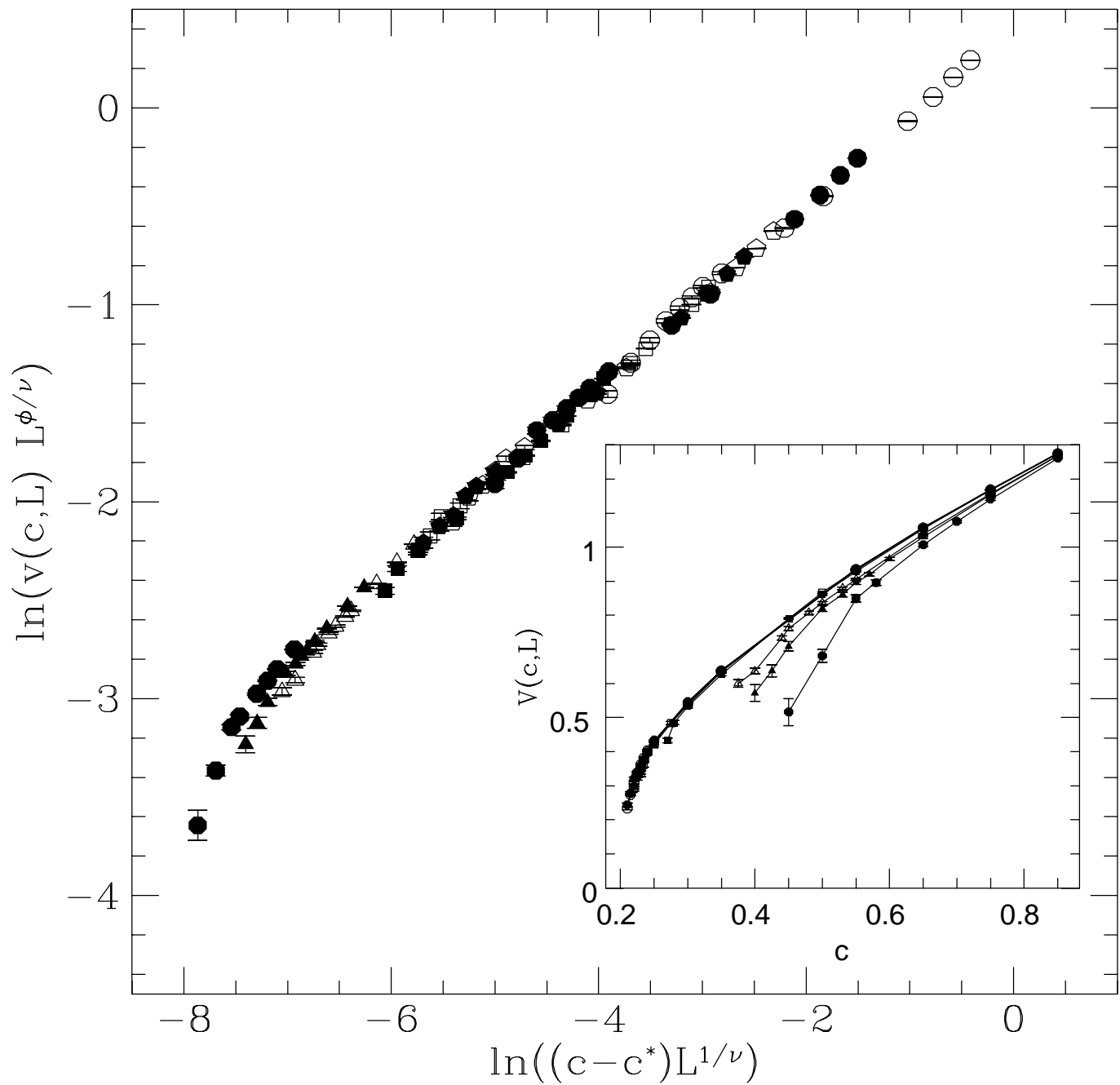
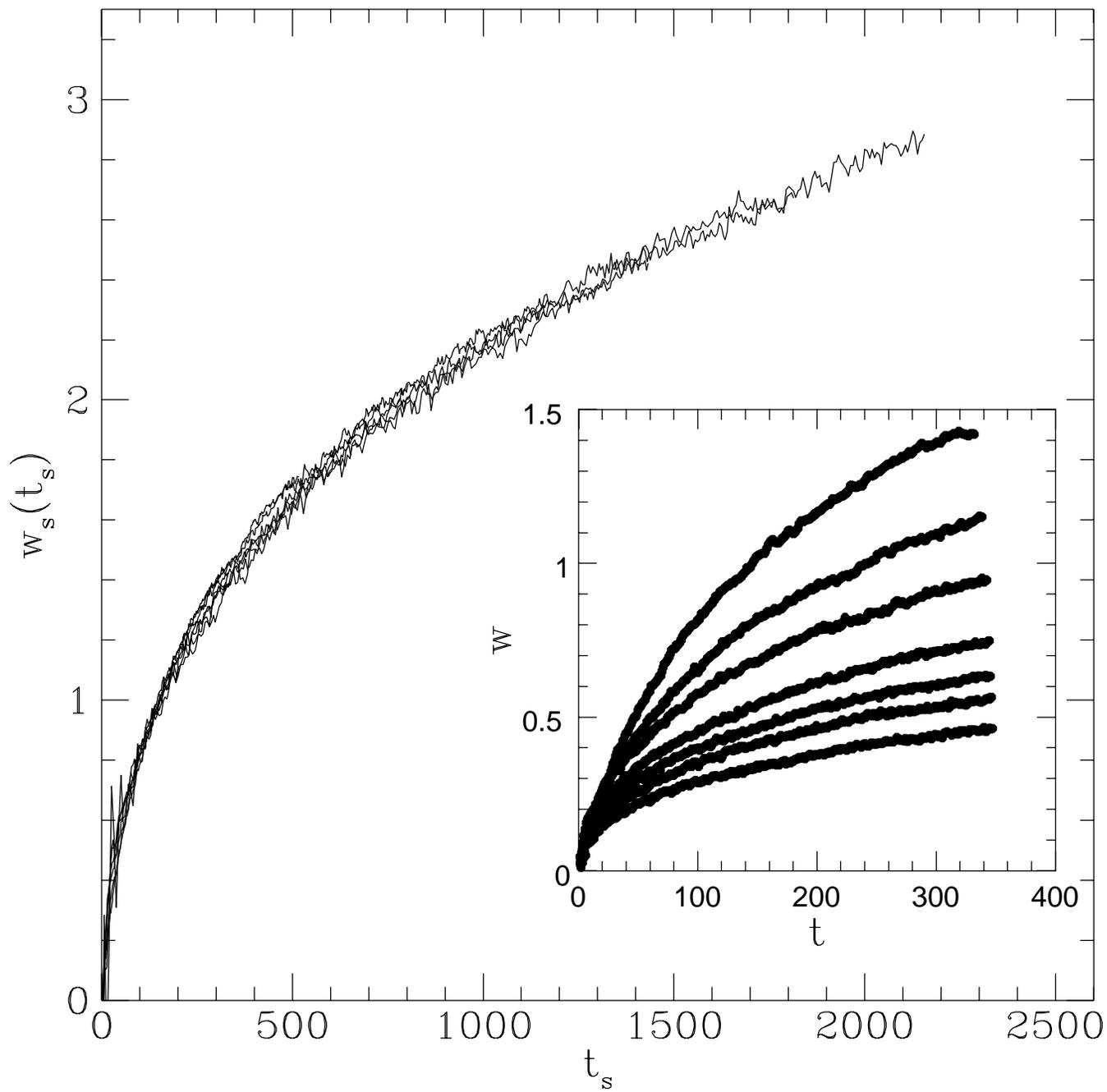
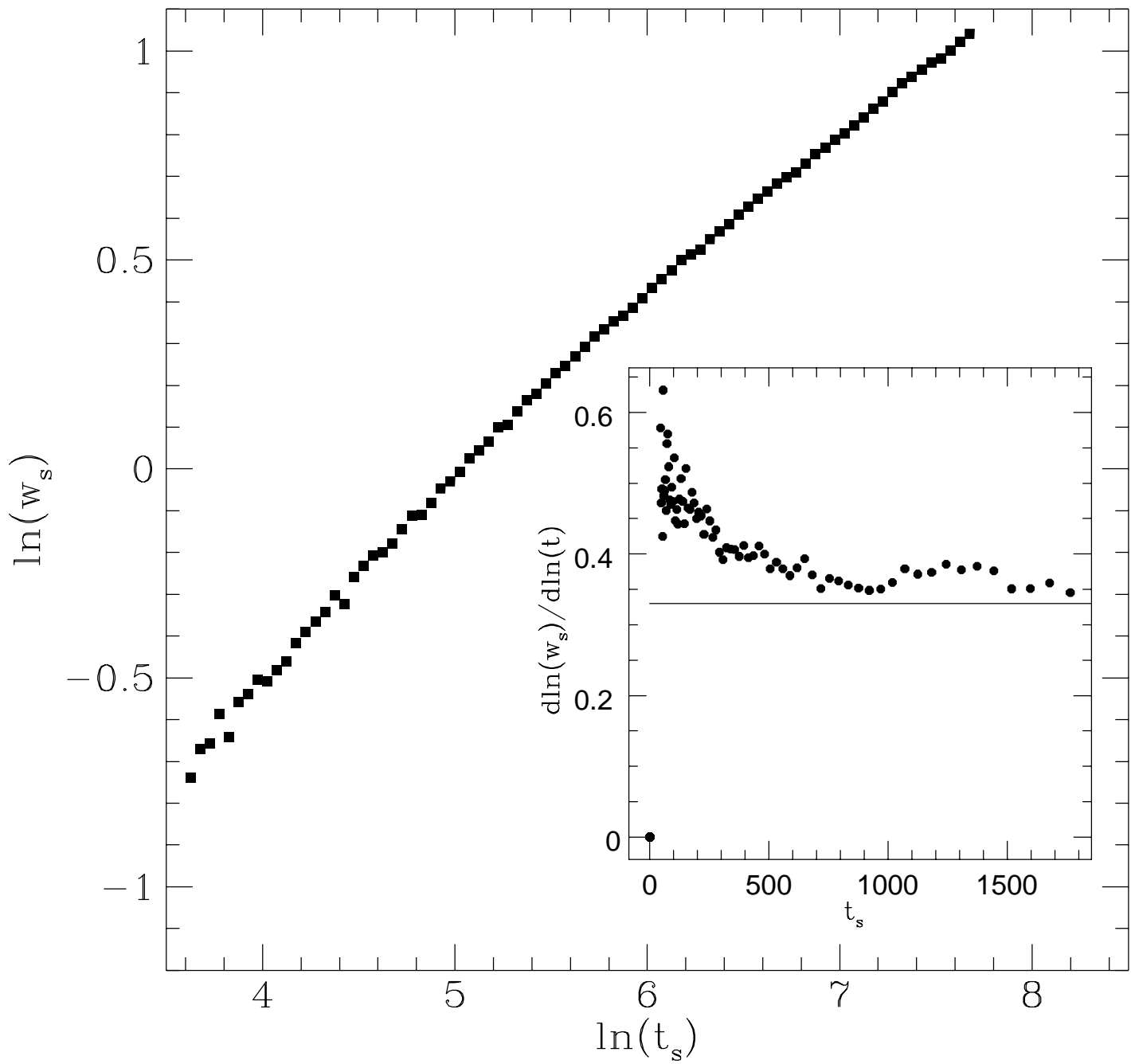
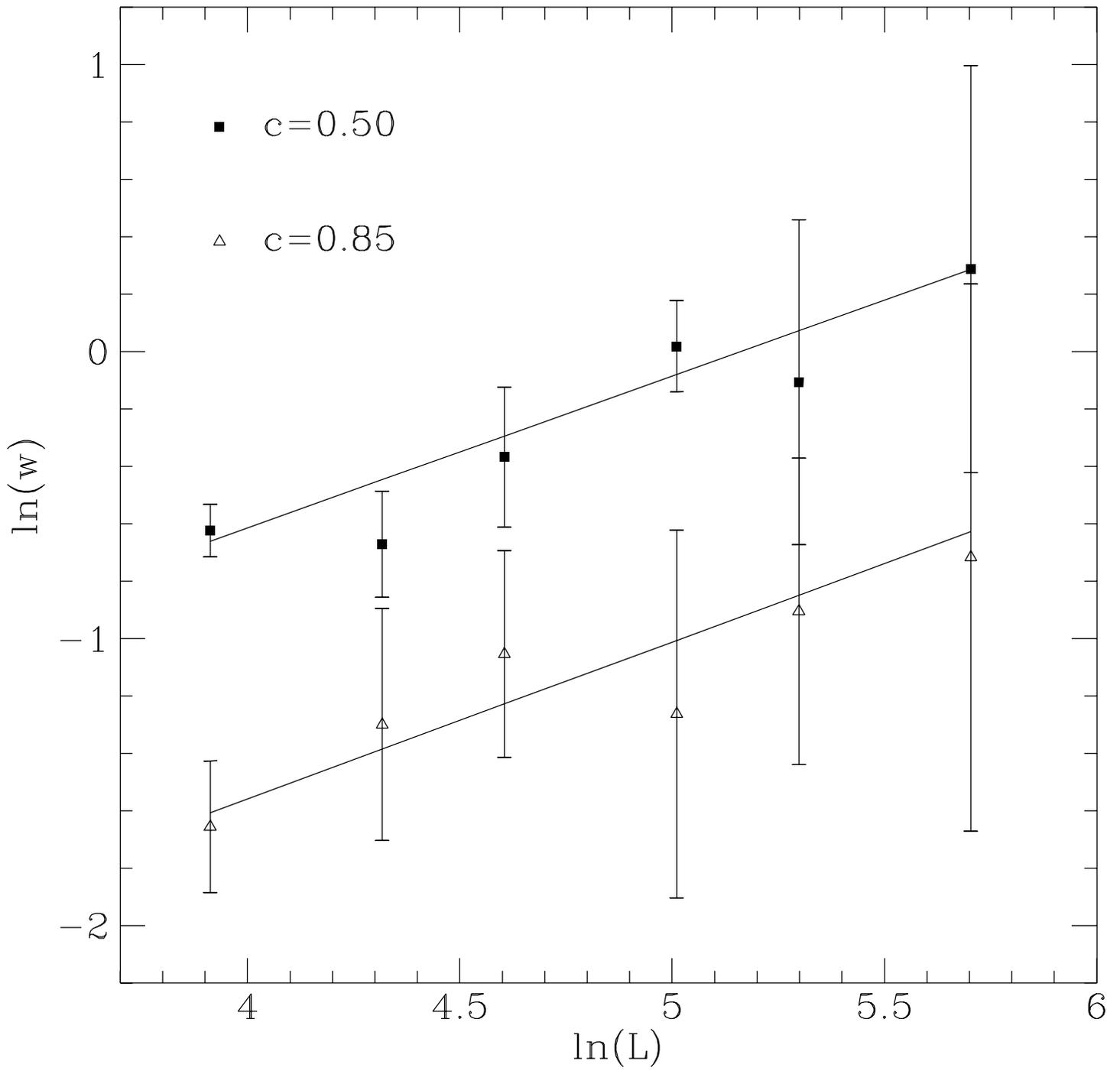
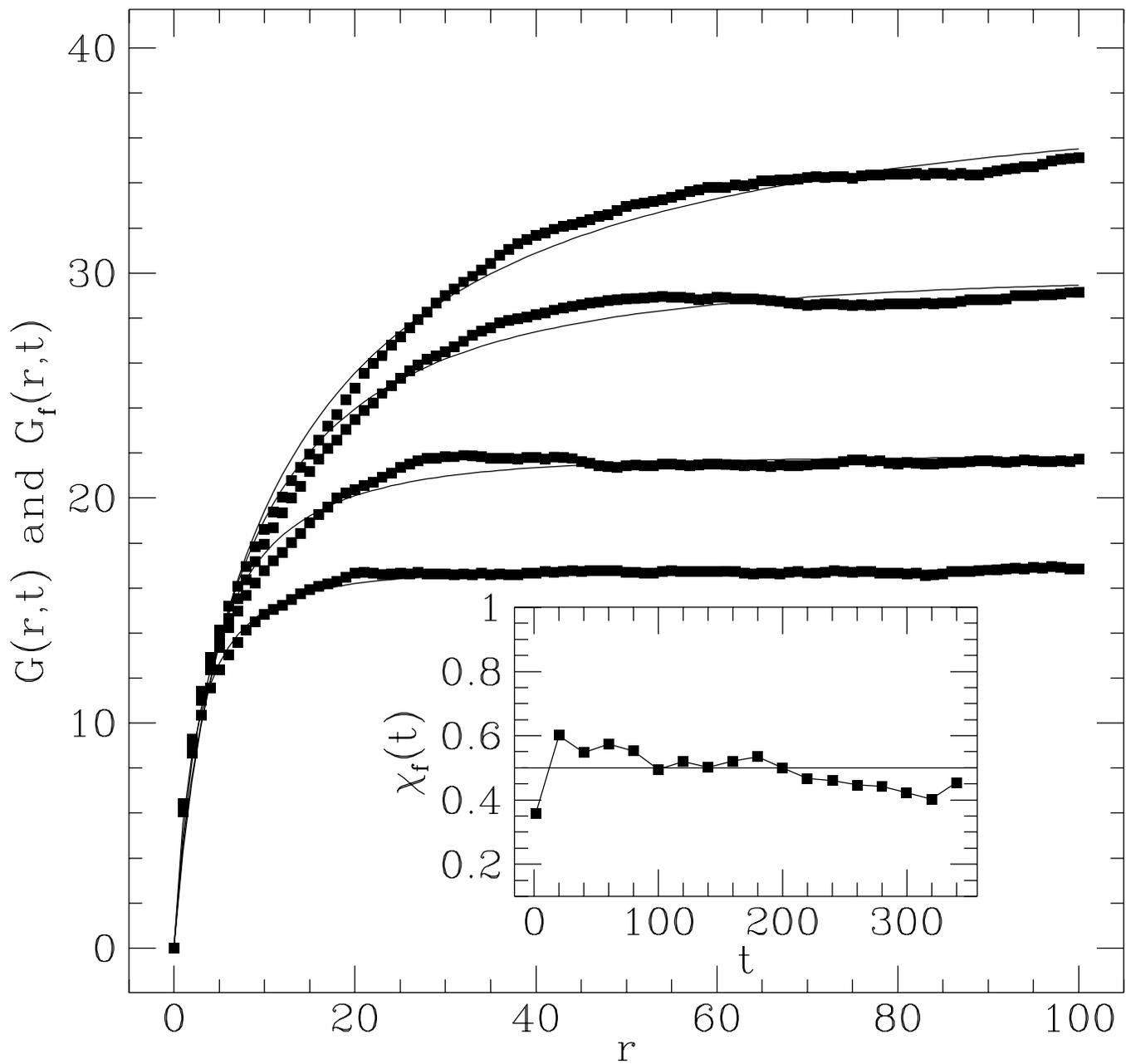
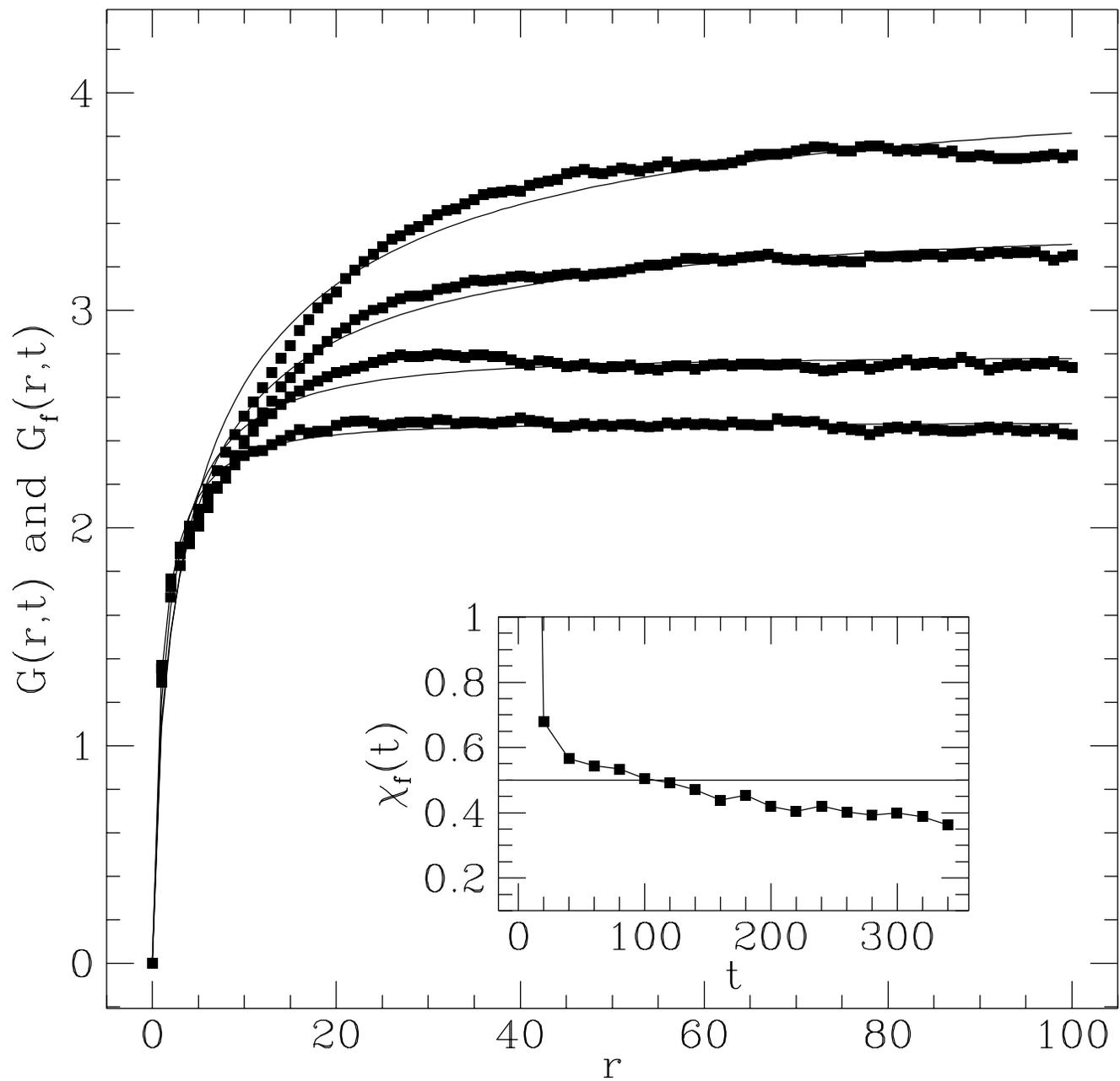
\end{document}